\pdfoutput=1
\documentclass[aps,pra,twocolumn,10pt]{revtex4-1}
\usepackage[utf8]{inputenc}
\usepackage{fourier}
\usepackage{amsmath,graphicx,microtype,bm}
\usepackage[colorlinks=true,citecolor=blue,urlcolor=blue]{hyperref}

\begin{document}

\title{Effect of short-range electron correlations in dynamic transport in a Luttinger liquid}

\author{Vladimir A.\ Sablikov and Yasha Gindikin}

\affiliation{Kotel'nikov Institute of Radio Engineering and Electronics, Russian Academy of Sciences, Fryazino, Moscow District, 141190, Russia}

\begin{abstract}
The density operator in the Luttinger model consists of two components, one of which describes long-wave fluctuations and the other is related to the rapid oscillations of the charge-density-wave (CDW) type, caused by short-range electron correlations. It is commonly believed that the conductance is determined by the long-wave component. The CDW component is considered only when an impurity is present. We investigate the contribution of this component to the dynamic density response of a Luttinger liquid free from impurities. We show that the conventional form of the CDW density operator does not conserve the number of particles in the system. We propose the corrected CDW density operator devoid of this shortcoming and calculate the dissipative conductance in the case when the one-dimensional conductor is locally disturbed by a conducting probe. The contribution of the CDW component to conductance is found to dominate over that of the long-wave component in the low-frequency regime.
\end{abstract}

\maketitle

\section{Introduction}

Electron transport in quantum wires is of significant interest because of the fundamental role that electron-electron interaction plays there. It is known that an arbitrarily small interaction destroys the quasiparticle excitations in a one-dimensional (1D) normal Fermi system~\cite{Hal,voit}. The excitations of an interacting 1D system are of bosonic nature. They represent the waves of electron density, transferring charge, and producing current. These waves have two components: one is smooth on the scale of the Fermi wavelength, and the other rapidly oscillates in space.

The contribution of the smooth density component to transport is now well investigated~\cite{cuniberti1998ac,PhysRevB.52.R5539,sablikov1998electron}. However, the theoretical predictions pertaining to this part of the density are not unambiguously supported by experiment to date~\cite{tarucha1995reduction,yacoby1996nonuniversal,thomas1998interaction}. Moreover, there is a principal question of what effects, connected inherently with electron-electron interaction, are observable in transport. Turning to this question, we draw attention to the fact that the Luttinger liquid is a strongly correlated electron state. Therefore, exploring interaction effects inevitably must involve the rapidly oscillating density component, which describes short-range electron correlations. 

This component of electron density comes from the interference of electrons, moving in opposite directions. It resembles the charge density waves (CDWs) in Peierls systems, but in contrast to them it is not accompanied by a lattice deformation~\cite{emery1979pp}. The characteristic wave number of the CDW component is $2k_F$, where $k_F$ is a Fermi wave number. 

The $2k_F$ density component is usually taken into account when the system contains an impurity, which pins the Luttinger liquid~\cite{kane1992transport,kane1992transmission,glazman1992quantum}. Then the electron density oscillates around the impurity, producing a soft gap in the electron density of states. As a consequence, the dc conductivity is suppressed. It is obvious that since short-range electron correlations contribute to the density, they must also contribute to transport even if any impurities are absent. 

The CDW susceptibility was considered in the pioneering work of Luther and Peschel~\cite{luther1974single}. In the present paper we investigate the CDW contribution to the dynamic density response function, susceptibility, and dissipative conductance of a single-mode quantum wire.

The first problem we meet here is that a conventional bosonized density operator describing the CDW does not conserve the number of particles in the system and, consequently, fails to describe correctly the 1D electron transport. We propose an expression for the density operator that is free from the mentioned shortcoming and calculate the CDW contribution to the density response function. The imaginary part of the CDW susceptibility exists in the narrow band of wave numbers near $2k_F$, whereas the smooth component produces a $\delta$-peak in the long-wave region. Both the CDW and smooth terms give, generally speaking, comparable contribution to the dissipative conductivity, which can be determined via the absorbed power. 

In order to estimate the CDW contribution to dissipative conductance, we consider the scheme of experiment, where a 1D conductor is subjected to the local external potential and the dissipated power is measured. This kind of experiment could be realized with the use of the probe microscopy technique, which allows one to produce the short-scale disturbance with a sufficiently large Fourier component at $2k_F$. Our calculations have led to the unexpected conclusion that the CDW contribution to the conductivity dominates in the low-frequency limit over the contribution due to the commonly discussed smooth component of the density.

\section{Density operator in the Luttinger model}

In this section we show that a standard form of the density operator describing the CDW is not consistent with the requirement of the particle number conservation. Hence, it is to be altered. We aim to obtain an adequate expression for the 1D electron density operator.

If the system is subjected to the external action that does
not change the total number of electrons, then the following
relation must be fulfilled at every moment of time t:
\begin{equation}
\label{int}
	\int^{L/2}_{-L/2}dx\,\langle\rho(x,t)\rangle = 0\,,
\end{equation}
where $\langle\rho(x,t)\rangle$ is a density fluctuation at the position $x$, and $L$ is the length of the system. This requirement can also be formulated for the susceptibility $\chi_{q \omega}$ of the system to external potential. The susceptibility $\chi_{q \omega}$, defined by the linear relation between the Fourier-transform of the density $\langle \rho \rangle_{q \omega}$ and the external potential $\varphi_{q \omega}$ via
\begin{equation}
\label{lim}
	\langle \rho \rangle_{q \omega} = \chi_{q\omega}\varphi_{q\omega}\,,
\end{equation}
must have the following limit when $\omega \ne 0$:
\begin{equation}
	\chi_{q \omega} \xrightarrow[q \to 0] {} 0 \,.
\end{equation}

Equations~\eqref{int} and~\eqref{lim} do not, of course, contradict the fact that an external charge (e.g.\ an impurity) is screened by 1D electrons. The external potential redistributes the electron density along the 1D conductor. The screening charge is concentrated near to the external charge, while the equal charge of the opposite sign goes to the wire ends. The screening charge is obviously determined by the limit of $\chi_{q \omega}$ at $\omega = 0$ when $q \to 0$. This limit is not equal to zero.

In the spinless Luttinger model, the electron density operator $\rho(x)$ is usually taken as~\cite{voit}
\begin{equation}
	\rho(x) = -\frac{1}{\pi} \partial_x \phi + \frac{k_F}{\pi}\cos(2 k_F x - 2 \phi)\,,
\label{Lutt_dens}
\end{equation}
where $\phi(x)$ is a bosonic phase. The first component of the density operator describes long-wave fluctuations, which are smooth on the $k_F$ scale and small as compared to background density $k_F/\pi$. This part of the electron density operator represents the sum of the densities of the right- and left-moving $r$-fermions (denoted by $r = \pm 1$). The second (CDW) component of the electron density operator is due to the interference of the left- and right-moving electrons.

It is easily seen that the form of the smooth density component $\rho_\mathrm{sm}(x)$ obeys the particle-number-conservation requirement. The integral of $\rho_\mathrm{sm}(x)$ is zero in consequence of boundary conditions at the ends of the 1D system:
\begin{equation}
	\int^{L/2}_{-L/2} dx \, \rho_\mathrm{sm}(x)=-\frac{1}{\pi}\phi(x)\bigg|^{+L/2}_{-L/2} = 0\,.
\end{equation}
The corresponding susceptibility $\chi_\mathrm{sm}(q,\omega)$ indeed goes to zero when $q \to 0$, as can be seen from the expression for $\chi_\mathrm{sm}(q,\omega)$ given in Ref.~\cite{cuniberti1998ac} (see also Eq.~\eqref{smsus} below). However, the CDW part of the electron density operator does not exhibit such a property. The integral of $\rho_\mathrm{CDW}(x)$ is not zero: 
\begin{equation}
	\int_{-L/2}^{L/2}dx\,\rho_\mathrm{CDW}(x)=\frac{k_F}{\pi}\int_{-L/2}^{L/2} dx\,\cos(2 k_F x - 2\phi ) \ne 0 \,.
\end{equation}

This implies that the integral of the average density $\langle\rho_\mathrm{CDW}(x)\rangle$ over $x$, generally speaking, is not zero too.  It is for this reason that the susceptibility $\chi_\mathrm{CDW}(q,\omega)$ calculated with the density operator~\eqref{Lutt_dens} does not go to zero when $q\to 0$, as we show in Section~\ref{SectCDW}. 

Consequently, the above expression for $\rho_\mathrm{CDW}$ and the corresponding density response function $\chi_\mathrm{CDW}$ contradict particle number conservation and do not correctly describe 1D electron transfer.

In order to understand the reasons of this failure, we ought to turn to the derivation of Eq.~\eqref{Lutt_dens} for the electron density operator in the Luttinger model.

The field operator $\Psi_\mathrm{el}(x)$ of electron system is known to be non-locally connected with that of $r$-fermion one $\Psi_r(x)$~\cite{voit}. However, in the \emph{low-energy} approximation it is usually taken as 
\begin{equation} 
	\Psi_\mathrm{el}(x)\simeq \Psi_+(x) + \Psi_-(x)\,. 
\end{equation} 
This local expression for a field operator leads to the electron density operator of Eq.~\eqref{Lutt_dens}, which does not conserve the number of particles.

The physical reason of the violation of the particle number conservation is that the density response due to the $2k_F$ part of the density of Eq.~\eqref{Lutt_dens} includes the response of the infinite sea of positrons, which is not completely eliminated by the normal ordering procedure.

Formally, the problem is that the consistent low-energy expansion of $\Psi_\mathrm{el}(x)$ has not been done. In order to find the proper form of the operator, one can either use the correct nonlocal expression for $\Psi_\mathrm{el}$ or try to find an adequate correction to the density operator of Eq.~\eqref{Lutt_dens}. We use the second option. Keeping in mind that the density operator is a model one, we bring it to the exact differential form by adding the lower order term in the low-energy expansion:
\begin{equation}
	\begin{split}
		\rho_\mathrm{CDW}(x)&= \frac{k_F}{\pi}\left(1-\frac{\partial_x \phi}{k_F}\right)\cos(2k_F x - 2\phi)\\ 
		&= \frac{1}{2\pi} \partial_x \sin(2k_F x - 2\phi)\,.
	\end{split}
	\label{rho_cdw}
\end{equation}
This form of the CDW operator guarantees the number of particles to be conserved. The integral of corrected $\rho_\mathrm{CDW}$ is zero in consequence of boundary conditions, and the susceptibility has a correct long-wave limit:  $\chi_\mathrm{CDW}(q,\omega) \sim q^2$ when $q\to 0$.

Another formula for the density operator, conserving the number of
particles, was proposed by Haldane~\cite{haldane1981effective}. He suggested to take a
function $\tilde \phi(x)$ that increases by $\pi$ on each encountered
electron. Then the electron density operator is presented as 
\begin{equation}
	\rho(x) = \frac{1}{\pi} \frac{\partial}{\partial x} \left(\tilde\phi + \sum_{m\ne 0} \frac{e^{2 i m \tilde\phi}}{2im} \right)\,.
\label{Halfor}
\end{equation}
If we associate $\tilde\phi(x)$ with $k_F x - \phi(x)$, then the first term of Eq.~\eqref{Halfor} gives the smooth part of the density in Eq.~\eqref{Lutt_dens}. The first harmonic ($m=\pm 1$) of Eq.~\eqref{Halfor} coincides with Eq.~\eqref{rho_cdw} in form, but differs by a factor of two.

In Section~\ref{SectCDW} we show that the density response, calculated with operator of Eq.~\eqref{rho_cdw}, gives the correct transition to the limiting case of noninteracting electrons. By contrast, Eq.~\eqref{Halfor} and the conventional expression~\eqref{Lutt_dens} for the CDW operator fail to give this transition.

Thus, the bosonized density operator~\eqref{rho_cdw} of the CDW component is consistent with the particle-number-conservation requirement and can be used to describe correctly the density response of interacting electrons.

\section{CDW contribution to transport}
In this section the contribution of the $2k_F$ and long-wave components of the density to the linear response function and to the dissipative conductance is explored. 

\label{SectCDW}
\begin{figure}[htb]
	\centering
	\includegraphics[width=0.9\linewidth]{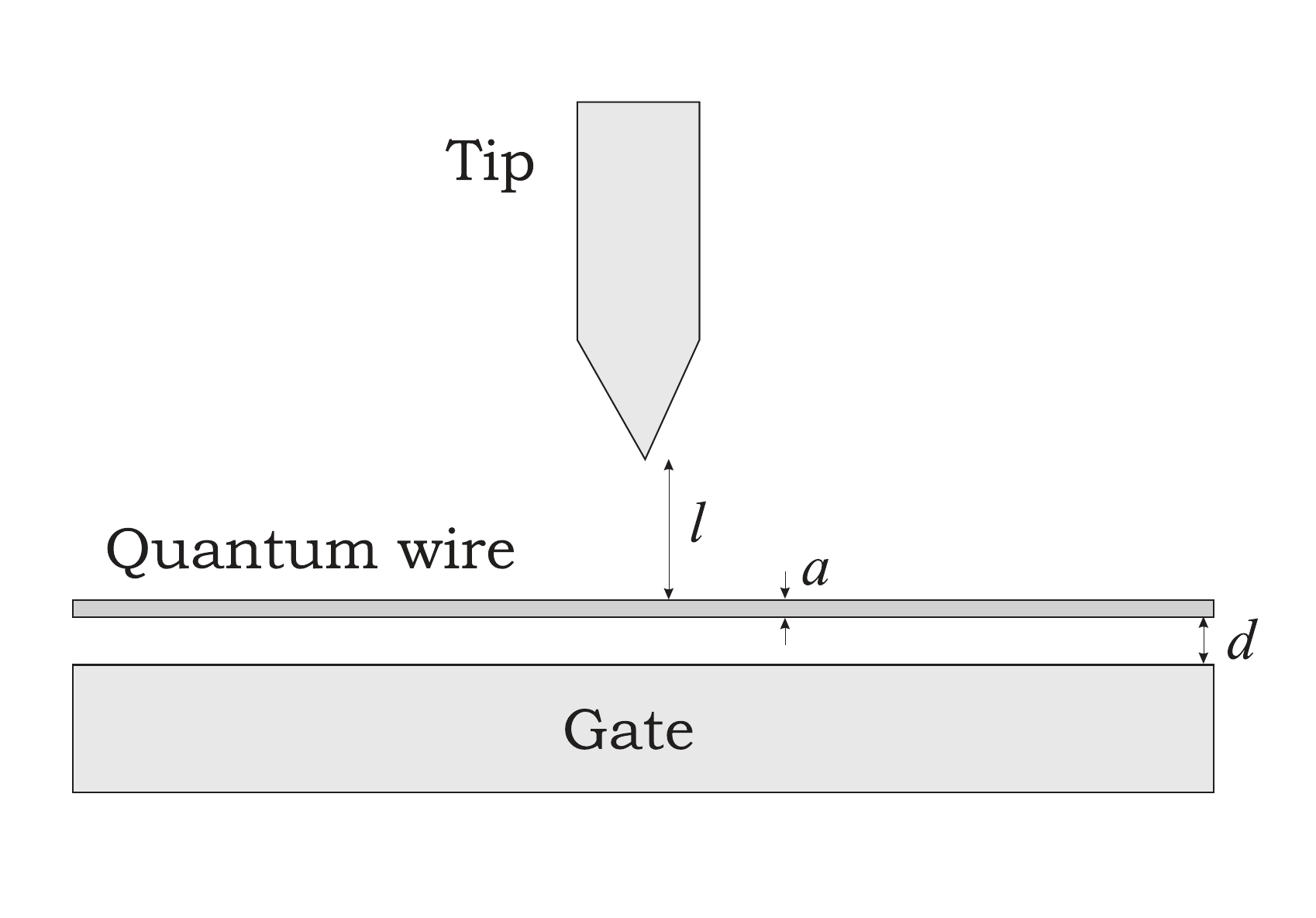}
	\caption{ A view of the model 1D system with short-range interaction. The tip produces a local potential disturbance.}
	\label{Pict1}
\end{figure}

To be specific, we consider the system, illustrated by Fig.~\ref{Pict1}, that comprises the gated single-mode quantum wire and the conducting tip. The tip allows one to act locally on the electrons in the wire by ac-potential $\varphi_\mathrm{ext}(x,t)$ that is localized in space: $\varphi_\mathrm{ext}(x \to \pm \infty) \to 0$. Such a configuration is very interesting in view of the rapid progress of the probe microscopy technique~\cite{PhysRevLett.83.4844,crook1999modification}.

We assume that the following hierarchy of the scales takes place:
\begin{equation}
	a \ll d \ll k_F^{-1} < l \,,
\end{equation}
 where $a$ is the diameter of the quantum wire, $d$ is the distance between the wire and the gate, and $l$ is the distance between the tip and the wire. The first pair of inequalities allows one to consider electrons in the wire as a Luttinger liquid with short-range interaction. The right inequality means that the influence of the tip is much weaker than that of the gate. Thus the presence of the tip does not affect electron-electron interaction and the ground state of 1D electrons is uniform.

The conductance of a mesoscopic structure is determined usually through the current and the voltage applied across the leads. Since no current-carrying leads are taken into our consideration, the conductance is determined by the dissipated power $P$ via $\sigma = P/V^2$, where $V$ is an external potential amplitude in the wire~\cite{cuniberti1998ac}.

If the electrons in the wire are subjected to the external electric potential of the form $\varphi_\mathrm{ext}(x,t) = \varphi(x)\cos \omega t$, the power dissipated in the system is shown in the Appendix to be
\begin{equation}
	P = -e^2 \omega \int^{+\infty}_{0} \frac{dq}{2\pi} |\varphi_q|^2\chi''(q,\omega)\,.
\label{pow}
\end{equation}

We calculate the density response function using the
Kubo formula and the density operator $\rho=\rho_{\rm sm}+\rho_\mathrm{CDW}$, with $\rho_\mathrm{CDW}$ being defined by Eq.~\eqref{rho_cdw}. The off-diagonal terms arising from $\langle \rho_\mathrm{sm}\rho_\mathrm{CDW} \rangle$ product do not contribute to the susceptibility in the thermodynamic limit $L \to \infty$. The contributions of the long-wave and CDW components are analyzed below.

\subsection{The long-wave density fluctuations}

The density response function, corresponding to $\rho_{\rm sm}$,
is well investigated~\cite{cuniberti1998ac}. Nevertheless it is given below to
demonstrate that the particle-number conservation is kept, and to
compare the dynamics of smooth density response with that of CDW:
\begin{equation}
	\chi_\mathrm{sm}(x,t)=
	\frac{g}{h} \theta(t)\,\partial_x[\delta(vt+x)-\delta(vt-x)]\,,
\label{smdyn}
\end{equation}
where $g$ is the interaction parameter and $v$ is the velocity of boson excitations~\cite{voit}. It is obvious that the integral of $\chi_\mathrm{sm}(x,t)$ over $x$ is zero at every moment of time.  The susceptibility
\begin{equation}
	\chi_\mathrm{sm}(q,\omega) = \frac{g}{h}\frac{2 q^2 v}{{(\omega+i0)}^2-q^2 v^2}\quad.
\label{smsus}
\end{equation}
goes to zero when $q\to 0$ and $\omega \ne 0$. This means that the number of particles is conserved.

Eq.~\eqref{smdyn} shows that the evolution of the electron density fluctuation generated by the local disturbance of the form $\varphi(x,t)\sim\delta(x)\delta(t)$ is presented by two wave fronts propagating in opposite directions. These waves transfer charge and produce current.

The fact that electron current depends on the position $x$ along the 1D conductor should not be perceived as a confusing one. It does not contradict to the general requirement of the electrodynamics according to which the total current $I$ must be conserved, i.e., be independent of $x$. The current conservation is obeyed if we take into account the displacement current. It was shown in Ref.~\cite{blanter1998interaction} that the displacement current between the 1D conductor and the gate ensures the total current conservation.

The dissipated power due to the smooth density component is
\begin{equation}
	P_\mathrm{sm} = g \frac{e^2}{2h} \frac{\omega^2}{v^2}|\varphi_{\omega/v}|^2 \,,
\label{sm_pow}
\end{equation}
where $\varphi_{\omega/v}$ is the Fourier-component of $\varphi(x)$ at $q = \omega/v$.

\subsection{\texorpdfstring{$\bm{2k_F}$}{2kF} density fluctuations}
The density response function due to the CDW component is
\begin{align}
		\chi_\mathrm{CDW}(x,t) = &\frac{\theta(t)}{2\pi h}\frac{\partial^2}{\partial
		x^2}\left({\left[\frac{\alpha^2}{\alpha^2+(vt+x)^2}\frac{\alpha^2}{\alpha^2+(vt-x)^2}\right]}^{g/2}\right.\notag \\
		&{}\times \left.\phantom{\bigg |} \sin\left(\pi g\theta_{\alpha}(v^2t^2-x^2)\right)\cos(2k_Fx)\right)\,,
\label{osn}
\end{align}
where $\alpha=k_F^{-1}$, and
\begin{equation}
	\pi \theta_{\alpha}(v^2t^2-x^2) = \arctan\left[\frac{vt+x}{\alpha}\right]+\arctan\left[\frac{vt-x}{\alpha}\right]\,.
\end{equation}
$\theta_{\alpha}(z)$ is a smoothed step function that goes to the Heaviside function $\theta(z)$ when $\alpha \to 0$. It describes the smoothed fronts moving with the velocity $v$.
\begin{figure}[t]
	\centering
	\includegraphics[width=0.9\linewidth]{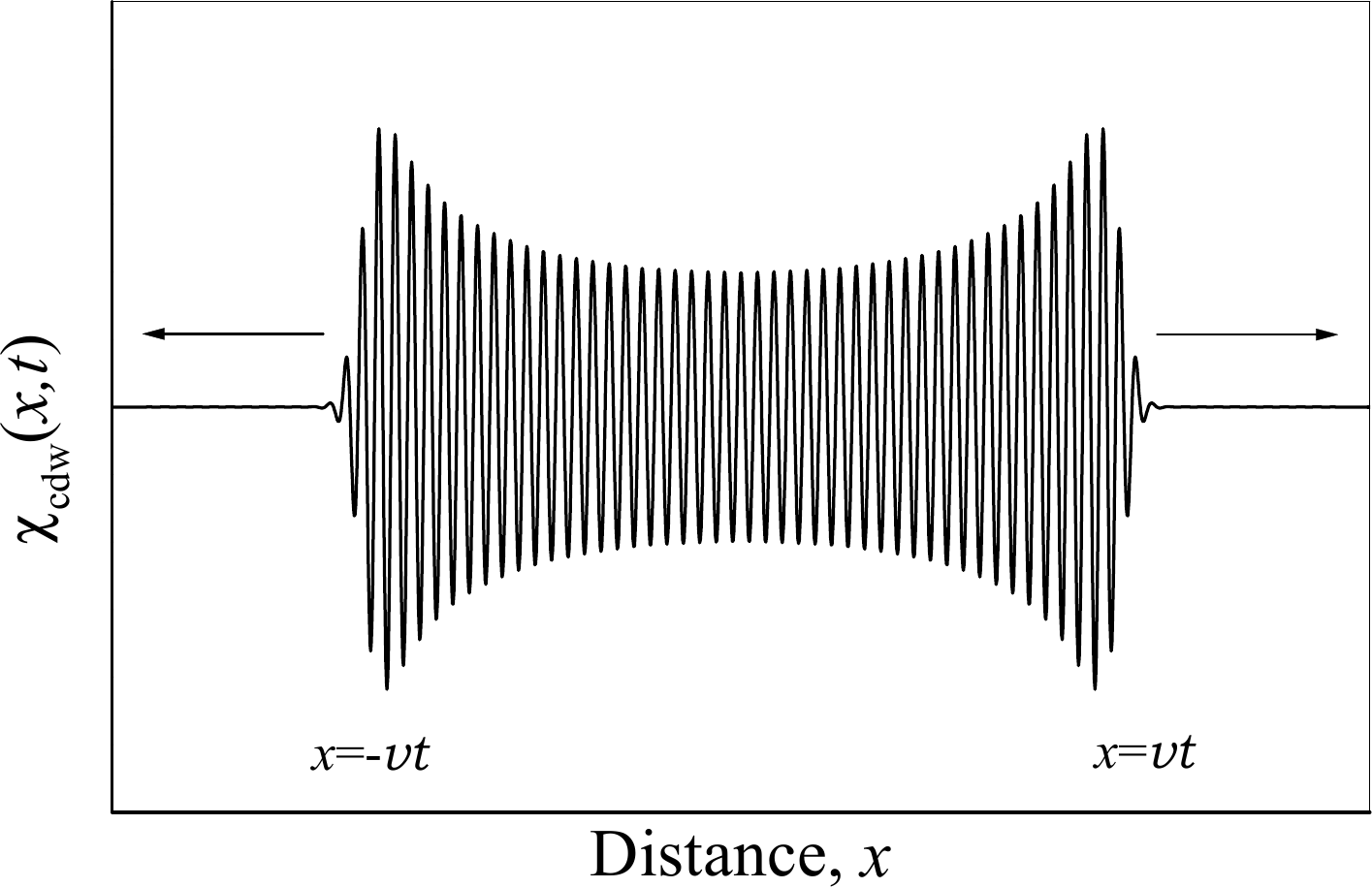}
	\caption{The CDW response function for a given $t$.}
	\label{Pict2}
\end{figure}

The integral of $\chi_\mathrm{CDW}(x,t)$ over $x$ is zero for any time $t$ because of the presence of the second derivative. The response function, calculated with the conventional expression~\eqref{Lutt_dens} for $\rho_\mathrm{CDW}$, differs from Eq.~\eqref{osn} just in the absence of the second derivative with respect to $x$.  The integral of this response function is obviously nonzero.

The response function $\chi_\mathrm{CDW}(x,t)$ represents the rapid oscillations modulated in the amplitude by the envelope that has the form of the moving wave, as illustrated in  Fig.~\ref{Pict2}. The wave front propagates with the velocity $v$, and the amplitude of the wave diminishes with time as $t^{-2g}$. The density oscillations reflect the physical picture of the disturbance in a Luttinger liquid. The evolution of the disturbance is determined by two kinds of motion. There exist compression waves, caused by forward scattering and described by Eq.~\eqref{smdyn}. In addition, the disturbed electrons suffer the backward scattering from the nearest particles, which gives rise to rapid density oscillations.

\begin{figure}[htb]
	\centering
	\includegraphics[width=0.9\linewidth]{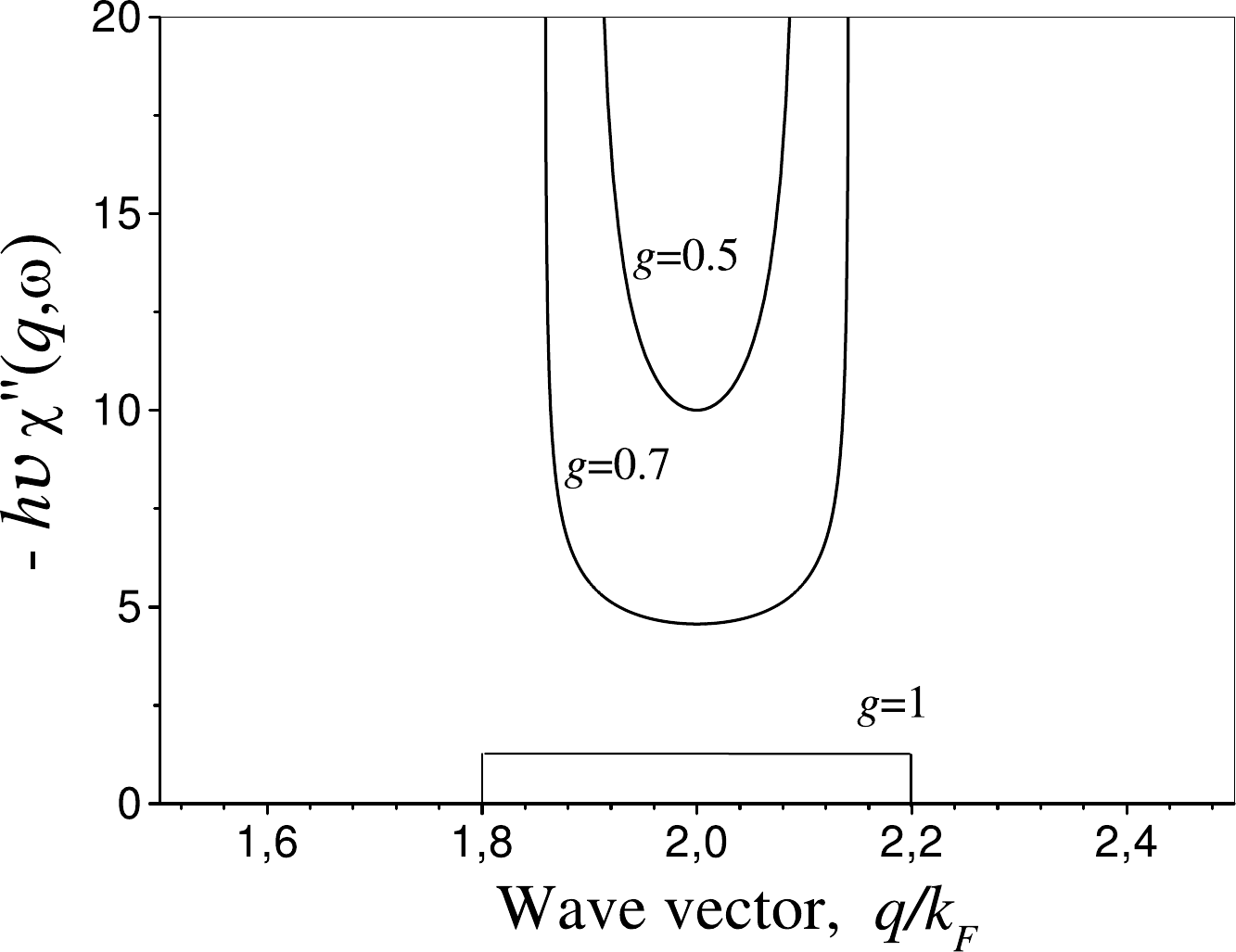}
	\caption{The imaginary part of the CDW susceptibility as a function of the wave number for the different values of the interaction parameter and $\hbar\omega/\varepsilon_F=0.2$.}
	\label{fig1}
\end{figure}

The CDW susceptibility equals
\begin{equation}
	\begin{split}
		\chi_\mathrm{CDW}(q,\omega) =& -\dfrac{1}{\hbar v} \dfrac{1}{4^{g+1}\Gamma^2(g)\sin(\pi g)}{\left(\dfrac{q}{k_F}\right)}^2\\
		&{}\times \sum\limits_{r=\pm 1} {\left[ {\left(\frac{q}{k_F}-2r\right)}^2 - {\left(\frac{\omega+i0}{k_Fv}\right)}^2 \right]}^{g-1}\,.
	\end{split}
\end{equation}

The imaginary part of the CDW susceptibility is nonzero only inside the interval $||q|-2k_F|<|\omega|/v$, where
\begin{equation}
	\chi_\mathrm{CDW}''(q,\omega) = -\dfrac{1}{\hbar v}\dfrac{{\rm
	sign}\,\omega}{4^g\Gamma^2(g)}{\left[{\left(\dfrac{\omega}{vk_F}\right)}^2-{\left(\dfrac{|q|}{k_F}-2\right)}^2\right]}^{g-1}\,.
\label{cdwsusc}
\end{equation}
Equation~(\ref{cdwsusc}) is valid for any interaction parameter $g$. When $g = 1$, Eq. (\ref{cdwsusc}) exactly reproduces the corresponding part of the Lindhard formula for free electrons. Incidentally, the susceptibility of free electrons, calculated with the density operator of Eq.~\eqref{Halfor}, differs from that given by the Lindhard formula by a factor.

The dependence of the imaginary part of the CDW susceptibility on the wave number $q$ is shown in Fig.~\ref{fig1} for several values of interaction parameter. It is seen that, in contrast to the noninteracting case, the inclusion of electron-electron interaction leads to the qualitative change in the shape of $\chi_\mathrm{CDW}''(q,\omega)$ and to the increase of $\chi_\mathrm{CDW}''(q,\omega)$ with interaction strength.

The real part of the susceptibility $\chi_\mathrm{CDW}'(q,\omega)$ is non-zero in the entire range of $q$. When $q$ goes to zero, $\chi_{\rm cdw}'(q,\omega)$ decreases as $q^2$, in accordance with the exact differential form of the expression~\eqref{rho_cdw} for the CDW density operator. The proportionality $\chi_\mathrm{CDW}'(q,\omega) \sim q^2$ takes place also for $\omega = 0$. Thus the exact differential form of the density operator not only guarantees the particle-number conservation, but also leads to the important conclusion that the CDW does not contribute to the screening charge.

The qualitative effect of the electron-electron interaction on the CDW susceptibility (on both the real and the imaginary parts) consists in the singularity that appears at $2k_F\pm \omega/v$. This singularity is obviously a dynamic analogue of Kohn's anomaly~\cite{luther1974single}. 

\begin{figure}[htb]
	\centering
	\includegraphics[width=0.9\linewidth]{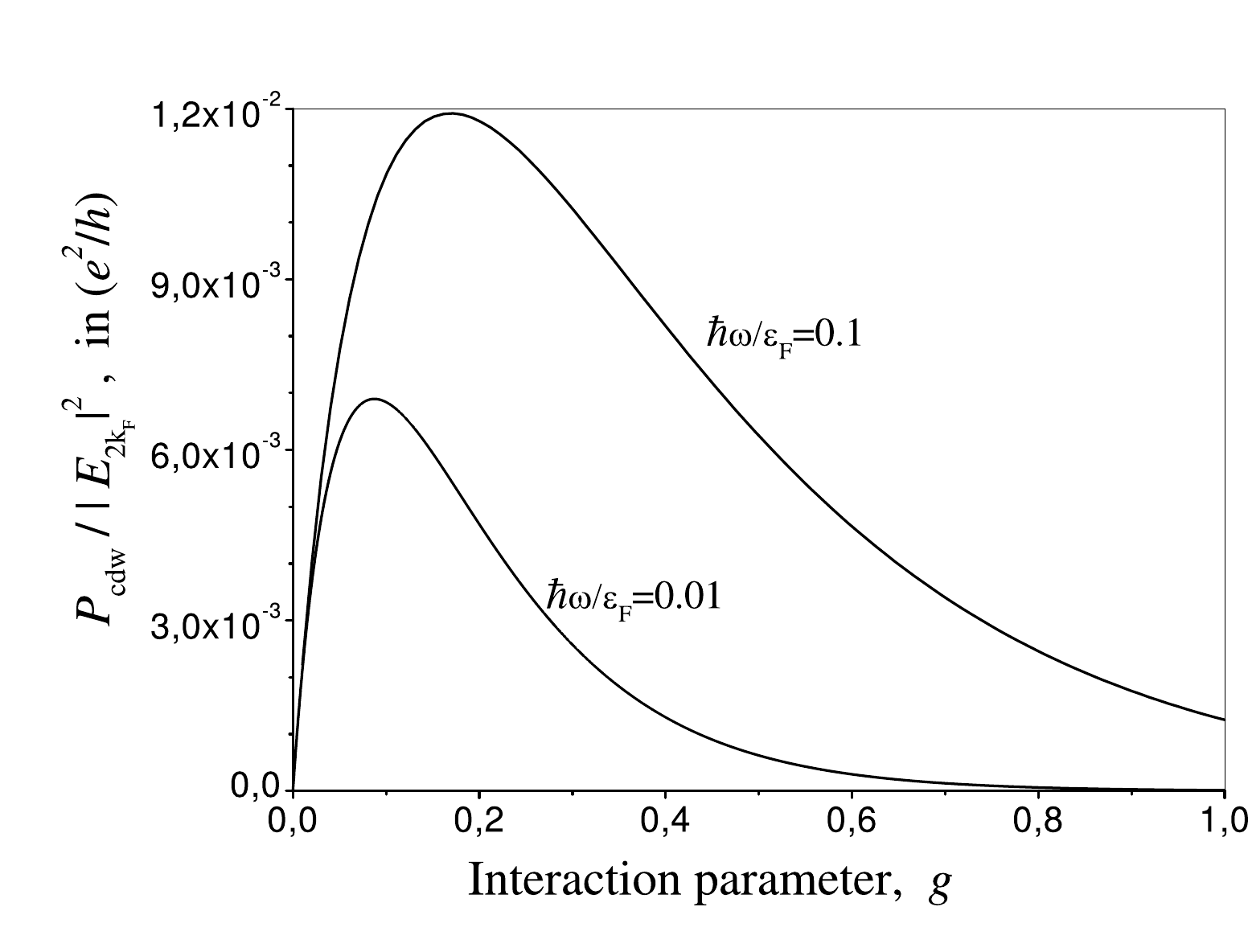}
	\caption{The dissipated power $P_\mathrm{CDW}$ as a function of the interaction parameter $g$. $E_{2k_F}$ is the Fourier-component of the external electric field at $q=2k_F$.} \label{pw}
\end{figure}

The dissipated power due to CDW is
\begin{equation}
	P_\mathrm{CDW}=\frac{e^2}{h}\frac{\Gamma(\frac12)}{4^g\Gamma(g)\Gamma(g+\frac12)}{\left(\frac{\omega}{vk_F} \right)}^{2g} k_F^2 \left|\varphi_{2k_F}\right|^2\,, 
\label{cdw_pow}
\end{equation}
where $\varphi_{2k_F}$ is the Fourier-component of the external potential at $q=2k_F$. The dependence of $P_\mathrm{CDW}$ on the interaction parameter $g$ is shown in Fig.~\ref{pw} for two frequencies. $P_\mathrm{CDW}$ is seen to have a pronounced peak at $g\approx {(2 \log(\varepsilon_F/\hbar\omega))}^{-1}$, where $\varepsilon_F$ is the Fermi energy. 

Let us compare the dissipated power for the CDW and conventional conductivity mechanisms. Equations~\eqref{sm_pow} and~\eqref{cdw_pow} show that
\begin{equation}
	\frac{P_\mathrm{CDW}}{P_{\rm sm}}\sim
	{\left(\dfrac{\omega}{vk_F}\right)}^{2g-2}\left|\dfrac{\varphi_{2k_F}}{\varphi_{\omega/v}}\right|^2\,.
\end{equation}
Since $g<1$, the CDW response is seen to dominate in the low-frequency limit.

The present analysis is valid for the low temperature $T < \hbar \omega$. The finite temperature effect can be taken into ac-
count in the same way as in Ref.~\cite{luther1974single}. Our calculations show that for $T > \hbar \omega$, the frequency dependence of the power becomes linear: $P_\mathrm{CDW} \sim \omega$. Thus the conclusion that the CDW mechanism of transport is the dominant one in the low-frequency regime holds for all temperatures and interaction parameters.

\section{ Conclusion}
In the present paper we have investigated the effect of short-range electron correlations on the dynamic conductivity of a 1D conductor within the frame of the Luttinger model. Short-range electron correlations are described by the CDW component of the bosonized density operator. We have found that a conventional form of this operator is not adequate to describe electron transfer. Namely, it does not conserve the number of particles in the system. We relate this inconsistency to the fact that the low-energy expansion of this operator is not properly performed in the Luttinger model. We have proposed a model operator of electron density that does not contradict to the total particle-number-conservation requirement and gives the correct transition to the case of noninteracting electrons.

The CDW density response function represents the rapid oscillations, modulated by the envelope function, which has the form of a moving wave. The CDW density component contributes significantly to the susceptibility of the 1D conductor without any impurities. The imaginary part of the CDW susceptibility is nonzero in the band of width $\omega/v$ around $2k_F$. Within this band the dissipative susceptibility depends strongly on the interaction parameter $g$. 

We have proposed a scheme of experiment in which the CDW mechanism of electron transport dominates over the commonly discussed mechanism due to the long-wave fluctuations. In this experiment a 1D conductor is disturbed locally by the ac potential applied to the conducting tip and the dissipated power is measured. In the low-frequency regime this power is determined predominantly by the CDW transport mechanism and drastically depends on the interaction parameter. The experiment discussed could give much more information about the nature of the correlated electron state.

\acknowledgments
This work was supported by INTAS (Grant No 96--0721), the Russian Fund for Basic Research (Grant No 99--02--18192), Russian Program ``Physics of Solid-State Nanostructures'' (Grant No 97--1054) and Russian Program ``Surface Atomic Structures'' (Grant No 5.3.99).

\section*{Appendix: The power dissipated in a gated wire}

The goal of this section is to obtain Eq.~\eqref{pow} for the power dissipated in the system shown in Fig.~\ref{Pict1}. When the voltage is applied to the tip, the external field induces charges in the wire and in the gate, so that the total electric potential is the sum of the external potential $\varphi_\mathrm{ext}(\bm{r},t)$ and the induced one $\varphi_\mathrm{ind}(\bm{r},t)$. The induced potential is determined by a linear equation
\begin{equation}
	\hat L\,\varphi_\mathrm{ind} = \rho(\bm{r},t)
\end{equation}
 where $\hat L$ is the Laplacian with appropriate boundary conditions at the gate surface. This equation coincides with the equation that governs the electron-electron interaction potential $ V(\bm{r})$: 
\begin{equation}
	\hat L\,V(\bm{r}-\bm{r'}) = -\delta(\bm{r}-\bm{r'})\,.
\end{equation}
Consequently, the interaction potential is a Green's function of this equation with boundary conditions imposed by the gate. So, $\varphi_\mathrm{ind}$ is connected with $V$ via
\begin{equation}
	\varphi_\mathrm{ind}(\bm{r})=-\int d\bm{r'}\,V(\bm{r}-\bm{r'})\rho(\bm{r'})\,.
\end{equation}

The effective 1D potential in the wire $\varphi(x,t)$ is connected with 3D potential $\varphi(\bm{r},t)$ through the following integral relation~\cite{sablikov1998coulomb} 
\begin{equation}
	\varphi(x,t) = \int d^2 r_{\perp}\,\varphi(\bm{r},t)\varrho(r_{\perp})\,,
\end{equation}
where $\varrho(r_{\perp})$ is a distribution function of electron density along the radial coordinate in the wire.

Hence, the induced potential in the wire satisfies the equation
\begin{equation}
	\varphi_\mathrm{ind}(q,\omega) = -U(q) \rho(q,\omega)\,,
\end{equation}
where $U(q)$ is a 1D effective interaction potential with account of the gate and the transverse density distribution in the wire. Thus the total potential and the external one are connected as follows
\begin{equation}
	\varphi_\mathrm{tot}(q,\omega) = \varphi_\mathrm{ext}(q,\omega) [1 - U(q)\chi(q,\omega)]\,.
	\label{fld}
\end{equation}
This relation, known for the homogeneous system, is seen to hold in a more complicated case of a 1D wire with a metal gate close by.

\vspace{1cm}
The dissipated power is given by
\begin{equation}
	P=\int_{-\infty}^{+\infty}dx\,\overline{E_\mathrm{tot}(x,t)j(x,t)}\,,
\end{equation}
where $j(x,t)$ is electron current and $E_\mathrm{tot}(x,t)$ is the total electric field; the overbar stands for time-averaging. Using Eq.~\eqref{fld} for the total potential and the continuity equation, one obtains the relation~\eqref{pow} for the dissipated power. The contribution of the induced potential to the power turns out to vanish during the time averaging, as shown in Ref.~\cite{cuniberti1998ac}.

The presence of the displacement current in the gap between the wire and the gate implies that electron current flows inside the gate. This current is of the same order of magnitude as the electron current in the wire. However, its contribution to the dissipated power is negligible because the conductivity of the gate is extremely high.

\bibliography{paper.bib}

\end{document}